\documentclass[pdflatex,sn-mathphys-num]{sn-jnl}

\usepackage{graphicx}%
\usepackage{multirow}%
\usepackage{amsmath,amssymb,amsfonts}%
\usepackage{amsthm}%
\usepackage{mathrsfs}%
\usepackage[title]{appendix}%
\usepackage{xcolor}%
\usepackage{textcomp}%
\usepackage{manyfoot}%
\usepackage{booktabs}%
\usepackage{algorithm}%
\usepackage{algorithmicx}%
\usepackage{algpseudocode}%
\usepackage{listings}%
\usepackage{tabularx}

\theoremstyle{thmstyleone}%
\theoremstyle{thmstyletwo}%

\theoremstyle{thmstylethree}%

\begin{document}

\title[Article Title]{Mathematical Modelling of Within-Host HIV Dynamics with Cytotoxic Immune Response and Antiretroviral Therapy}


\author*[1]{\fnm{Maxim} \sur{Polyakov}}\email{m.v.polyakov@volsu.ru}

\author[1]{\fnm{Yulia} \sur{Chernikova}}\email{prim-261\_657658@volsu.ru}
\equalcont{These authors contributed equally to this work.}


\affil*[1]{\orgdiv{Department of Information Systems and Computer Modelling}, \orgname{Volgograd State University}, \orgaddress{\street{ Universitetsky pr., 100}, \city{Volgograd}, \postcode{400062}, \state{Volgograd Region}, \country{Russia}}}




\abstract{We develop and analyse a mechanistic ODE model of within-host HIV dynamics that includes uninfected CD4$^+$ T cells, latently and productively infected cells, free virions, and cytotoxic immune effectors. Antiretroviral therapy is described by two time-dependent efficacy functions that separately reduce new infections and virion production. We show that solutions remain non-negative and bounded, identify the infection-free and endemic equilibria, and derive the basic reproduction number together with the local stability condition for the infection-free state. We also obtain treatment-dependent suppression thresholds and an analytical estimate of when this threshold is crossed as treatment efficacy declines. Numerical simulations consider untreated infection, ART initiation, periodic variation in efficacy, and progressive loss of treatment effect. Without therapy, the model approaches a state of persistent infection. ART reduces viral load and promotes recovery of the CD4$^+$ T-cell population, while the latent reservoir persists under the parameter set considered. Periodic changes in treatment efficacy produce sustained forced oscillations. These results connect the analytical threshold conditions with the treatment-dependent dynamics of the model and provide a basis for future calibration, sensitivity analysis, uncertainty quantification, and extensions informed by clinical data.}




\maketitle

\section{Introduction}\label{sec1}

Human immunodeficiency virus (HIV) remains a major global public health challenge despite substantial advances in diagnosis, prevention, and treatment. Current estimates indicate that tens of millions of people are living with HIV globally, while the annual numbers of new infections and deaths associated with acquired immunodeficiency syndrome (AIDS) continue to constitute a major medical and public health burden \cite{UNAIDS2025,WHO2025HIVStats,WHO2024HIVData}. The widespread implementation of combination antiretroviral therapy (ART) has fundamentally altered the prognosis of HIV infection. Provided that treatment is initiated in a timely manner and maintained consistently, HIV infection has evolved from a rapidly progressive and frequently fatal condition into a manageable chronic disease \cite{START2015,Gandhi2025,DHHS2025,Bekker2023}. Nevertheless, a broadly applicable curative treatment remains unavailable because ART suppresses active viral replication without eliminating all cellular and tissue sources of persistent HIV infection \cite{Chou2024,McMyn2023,Pasternak2023}.

A defining feature of HIV pathogenesis is the targeting of CD4$^+$ T lymphocytes, a central component of adaptive immunity. Viral replication, the death of infected cells, chronic antigenic stimulation, and dysregulation of immune responses collectively lead to a progressive decline in the concentration of functionally active CD4$^+$ T cells, thereby increasing susceptibility to opportunistic infections and malignancies \cite{Deeks2015,Bekker2023}. At the same time, HIV infection induces both innate and adaptive immune responses, including humoral mechanisms and cell-mediated cytotoxic responses driven primarily by CD8$^+$ cytotoxic T lymphocytes \cite{PerdomoCelis2019,Arenas2023,Papadopoulos2025}. Consequently, HIV dynamics cannot be adequately described solely in terms of interactions between free virions and target cells; the regulatory effects of the immune response on infected cells and viral load must also be taken into account.

The cytotoxic immune response plays a dual role in the course of HIV infection. On the one hand, CD8$^+$ T cells can recognise and eliminate productively infected cells, thereby limiting viral expansion, particularly during the early stages of infection and in certain groups of HIV controllers \cite{NowakBangham1996,PerdomoCelis2019,Arenas2023,Papadopoulos2025}. On the other hand, chronic antigenic stimulation, immune exhaustion, viral variability, and immune escape progressively impair the effectiveness of cytotoxic control and contribute to the establishment of persistent chronic infection \cite{Ganusov2013,Deng2021,Deng2023}. Incorporating the cytotoxic immune response into mathematical models of HIV infection is therefore essential for analysing the conditions under which the immune system can partially restrain viral replication without achieving complete viral elimination.

Antiretroviral therapy targets multiple stages of the HIV life cycle, including viral entry, reverse transcription, integration, virion maturation, and other processes \cite{Gandhi2025,DHHS2025,Kemnic2022}. Clinical studies have demonstrated that early treatment initiation improves individual outcomes and reduces the risk of viral transmission, while sustained virological suppression reduces the risk of sexual HIV transmission to negligible levels \cite{START2015,Cohen2016,Rodger2019,Eisinger2019}. However, therapeutic efficacy is determined not only by the pharmacological properties of antiretroviral agents but also by treatment adherence, drug resistance, pharmacokinetic fluctuations in drug concentrations, and patient-specific factors \cite{WHO2024DrugResistance,Manalel2024,Meiners2023,Kagan2025}. Time-dependent treatment efficacy functions can represent treatment initiation, irregular medication intake, and the gradual loss of efficacy associated with drug resistance.

One of the principal barriers to HIV eradication is the latent reservoir, which is established predominantly within long-lived populations of CD4$^+$ T cells. Early studies demonstrated that replication-competent virus can persist in resting CD4$^+$ T cells even during otherwise effective therapy \cite{Finzi1997,Chun1997,Siliciano2003}. Subsequent investigations further characterised the reservoir by identifying contributions from non-induced proviruses, clonal expansion of infected cells, T cells with stem-cell-like memory properties, and tissue compartmentalisation of viral persistence \cite{Ho2013,Buzon2014,Maldarelli2014,Wagner2014,Cohn2015,LorenzoRedondo2016}. Contemporary reviews and experimental studies confirm that latent infection and long-term reservoir persistence remain central obstacles to achieving either a functional or sterilising cure for HIV \cite{Sengupta2018,Chen2022,McMyn2023,Pasternak2023,Chou2024,Pieren2024}. A mathematical model designed to investigate within-host HIV dynamics should therefore include not only productively infected cells but also a population of latently infected cells.

Mathematical modelling has played a central role in establishing a quantitative framework for the study of HIV infection. Classical studies of viral dynamics enabled estimation of virion clearance rates, the lifespan of productively infected cells, and the turnover rate of the viral population \cite{Ho1995,Wei1995,Perelson1996,Stafford2000}. These findings provided the foundation for within-host models in which the dynamics of free virus, target cells, and infected cell populations are represented by systems of ordinary differential equations (ODEs) \cite{Perelson2013,Hill2018Treatment}. Such models were subsequently extended to incorporate ART, pharmacodynamics, treatment adherence, drug resistance, latent reservoirs, and immune responses \cite{Xiao2013,Hill2018Latency,DOrso2023,Vemparala2024,Rasi2025}. As a result, mathematical models have become useful not only for explaining observed viral-load dynamics but also for evaluating hypothetical therapeutic strategies that are difficult or impossible to investigate comprehensively in experimental settings.

Contemporary mathematical biology of HIV infection is increasingly moving beyond basic ``virus--target cell'' models towards more detailed systems that incorporate immune regulation, latency, time delays, infected-cell age structure, cell-to-cell transmission, spatial heterogeneity, and optimal therapeutic control \cite{Conway2015,RongPerelson2009,Hill2014,Bai2021,Guo2020,Liu2020}. Recent studies have also examined models incorporating cytotoxic immune responses, antibody-mediated responses, immune exhaustion, distributed delays, and different ART regimens, including treatment switching and the emergence of drug resistance \cite{Deng2021,Deng2023,Chen2024,AlShamrani2024,Viriyapong2024,Li2025,Nali2025}. Particular attention has also been paid to parameter identifiability, model-to-data comparison, equilibrium stability, and the conditions required for post-treatment viral control \cite{Conway2015,Liyanage2024,Rodriguez2025,Cai2025}. 

Despite the large number of existing HIV models, there remains a need for compact systems that retain biological interpretability while incorporating several key mechanisms of infection within a unified framework. Overly simplified models may fail to represent the contribution of latently infected cells and the cytotoxic immune response, whereas highly detailed models typically introduce large numbers of parameters that are difficult to identify and may substantially complicate analytical investigation \cite{Perelson2013,Hill2018Latency,DOrso2023,Liyanage2024}. An important objective is therefore to develop an intermediate-complexity model that preserves a clear biological interpretation of its state variables and parameters while accounting for uninfected CD4$^+$ T cells, latently infected cells, productively infected cells, free virions, cytotoxic immune effector cells, and the time-dependent effects of ART.

The aim of this study is to develop and analyse an interpretable within-host HIV model incorporating latent infection, cytotoxic immune control, and time-dependent antiretroviral therapy.

The model extends the classical within-host HIV framework in three directions: 
it includes a latent reservoir and a cytotoxic effector compartment, and it 
represents ART through separate time-dependent effects on cellular infection 
and virion production. Related mathematical models have addressed latent 
infection and post-treatment viral control 
\cite{Conway2015,Vemparala2024,Rasi2025}, cytotoxic immune responses 
\cite{Deng2021,Deng2023,Li2025}, and treatment-dependent HIV dynamics 
\cite{Xiao2013,Deng2023}. The resulting system remains sufficiently compact 
for analytical treatment. We derive the viral invasion threshold and its 
dependence on therapy and, for exponentially declining treatment efficacy, 
obtain an explicit estimate of the time at which the suppression threshold is 
crossed. Numerical simulations are then used to relate these analytical results 
to untreated infection, ART initiation, periodic variation in efficacy, and 
progressive loss of treatment effect.

The remainder of this paper is structured as follows. Section \ref{sec2} introduces the mathematical model, describes the numerical methods, and specifies the parameterisation and treatment scenarios considered in the simulations. Section \ref{sec3} presents the mathematical analysis of the model, including positivity and boundedness of solutions, infection-free and endemic equilibria, the basic reproduction number, and treatment-dependent threshold conditions, together with a numerical illustration of the loss of viral control under declining treatment efficacy. Section \ref{sec4} reports the results of the computational experiments for untreated infection, ART initiation, and periodically varying treatment efficacy. Section \ref{sec5} discusses the biological and mathematical interpretation of the obtained results, as well as the main limitations of the model. Finally, Section \ref{sec6} summarises the main conclusions and outlines directions for further development of the proposed framework.

\section{Model Formulation and Computational Methods}\label{sec2}

\subsection{Mathematical Model}
\label{subsec:model-formulation}

We consider a within-host model of HIV infection dynamics describing the interactions among free virions, CD4$^+$ T lymphocytes, infected cell populations, and the effector immune response. The model builds on the classical ``virus--target cell'' framework widely used to describe HIV dynamics \cite{Ho1995,Wei1995,Perelson1996,Perelson2013} and extends it by explicitly incorporating latently infected cells, productively infected cells, the effector immune response, and the time-dependent effects of antiretroviral therapy \cite{RongPerelson2009,Hill2018Latency,Hill2018Treatment,Xiao2013,PerdomoCelis2019}.

The model comprises five state variables, $T(t)$, $L(t)$, $I(t)$, $V(t)$, and $C(t)$, where $t$ denotes time measured in days. Their biological interpretations are summarised in Table~\ref{tab:variables}. All state variables are assumed to be non-negative functions of time.

\begin{table}[htbp]
\centering
\caption{State variables of the within-host HIV dynamics model.}
\label{tab:variables}
\begin{tabular}{ll}
\hline
Symbol & Biological interpretation \\
\hline
$T(t)$ & uninfected CD4$^+$ T cells \\
$L(t)$ & latently infected CD4$^+$ T cells \\
$I(t)$ & productively infected CD4$^+$ T cells \\
$V(t)$ & free HIV virions \\
$C(t)$ & immune effector cells \\
\hline
\end{tabular}
\end{table}

\begin{figure}[htbp]
\centering
\includegraphics[width=0.85\textwidth]{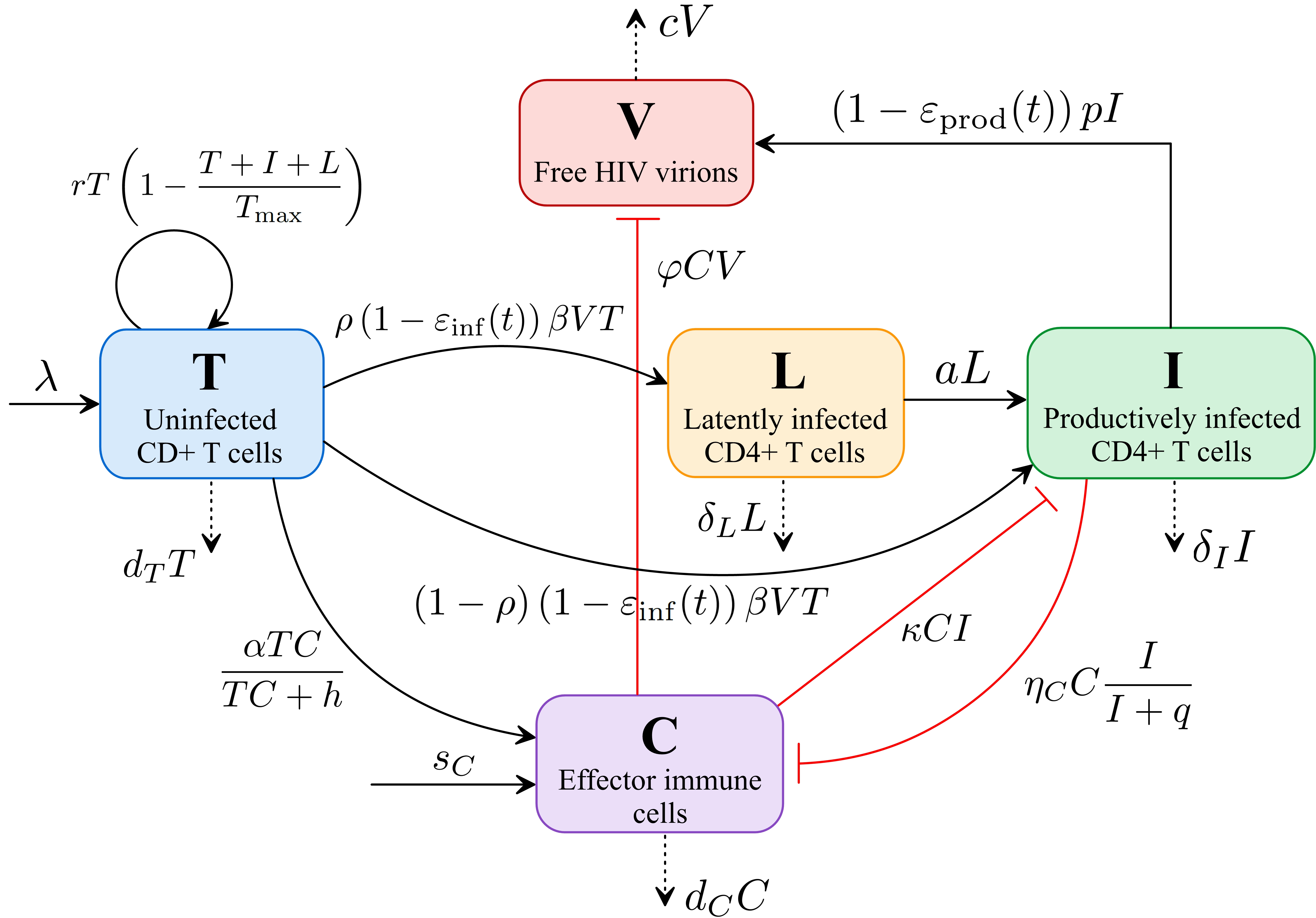}
\caption{Conceptual representation of HIV--immune system interactions underlying the mathematical model.}
\label{fig:model-scheme}
\end{figure}

A conceptual representation of the interactions among the principal model components is presented in Fig.~\ref{fig:model-scheme}. Free virions $V$ infect uninfected CD4$^+$ T cells $T$. Following infection, a fraction of the cells enters the latent state $L$, whereas the remaining fraction becomes productively infected and enters state $I$. Latently infected cells can reactivate and replenish the population of productively infected cells, which in turn produces new virions. Effector cells $C$ restrict infection by eliminating productively infected cells and mediating an additional reduction in the concentration of free virus. At the same time, prolonged antigenic stimulation may result in functional exhaustion of the effector response.

The biological structure of the model is based on the following assumptions. Uninfected CD4$^+$ T cells enter the system at a constant rate and are additionally maintained through homeostatic proliferation constrained by an effective carrying capacity. Infection of susceptible cells by free virions is described using mass-action kinetics. Following infection, a fraction of cells enters a latent state and does not directly produce virus, whereas the remaining cells become productively infected. Latently infected cells can subsequently reactivate. Productively infected cells generate new virions and are eliminated through natural death and cytotoxic killing by effector cells. Free virions are removed through natural clearance and immune-mediated viral clearance. Effector cells are maintained by a basal influx, are stimulated by the CD4$^+$ T-cell compartment, and may lose functional activity under prolonged antigenic stimulation.

Taking these processes into account, HIV infection dynamics are described by the following system of ordinary differential equations:
\begin{subequations}\label{eq:model}
\begin{align}
\frac{dT}{dt}
&=
\lambda
+rT\left(1-\frac{T+I+L}{T_{\max}}\right)
-d_TT
-\left(1-\varepsilon_{\mathrm{inf}}(t)\right)\beta VT,
\label{eq:model_T}
\\
\frac{dL}{dt}
&=
\rho\left(1-\varepsilon_{\mathrm{inf}}(t)\right)\beta VT
-aL
-\delta_L L,
\label{eq:model_L}
\\
\frac{dI}{dt}
&=
(1-\rho)\left(1-\varepsilon_{\mathrm{inf}}(t)\right)\beta VT
+aL
-\delta_I I
-\kappa CI,
\label{eq:model_I}
\\
\frac{dV}{dt}
&=
\left(1-\varepsilon_{\mathrm{prod}}(t)\right)pI
-cV
-\varphi CV,
\label{eq:model_V}
\\
\frac{dC}{dt}
&=
s_C
+\frac{\alpha TC}{TC+h}
-d_CC
-\eta_C C\frac{I}{I+q}.
\label{eq:model_C}
\end{align}
\end{subequations}

Equation~\eqref{eq:model_T} describes the dynamics of uninfected CD4$^+$ T cells. The parameter $\lambda$ represents their constant influx, $r$ is the rate of homeostatic proliferation, and $T_{\max}$ denotes the effective carrying capacity of the cell population. Natural cell loss is represented by the term $d_TT$. Infection of cells by free virions is described by the term $\beta VT$, where $\beta$ is the infection rate coefficient. The function $\varepsilon_{\mathrm{inf}}(t)\in[0,1]$ represents the efficacy of therapy in reducing the formation of newly infected cells.

Equation~\eqref{eq:model_L} describes the dynamics of latently infected cells. A fraction $\rho\in[0,1]$ of newly infected CD4$^+$ T cells enters the latent state. These cells do not directly produce virus but can reactivate at rate $a$ and subsequently enter the productively infected state. Natural loss of latently infected cells is governed by the parameter $\delta_L$. Inclusion of the latent compartment enables the model to represent the persistence of HIV within long-lived cellular reservoirs even when active viral replication is effectively suppressed \cite{Finzi1997,Chun1997,Siliciano2003,Ho2013,McMyn2023}.

Equation~\eqref{eq:model_I} describes the population of productively infected cells. These cells arise both from the direct transition of a fraction $1-\rho$ of newly infected cells into the productive state and from reactivation of latently infected cells. Loss of this population occurs through natural cell death at rate $\delta_I$ and cytotoxic killing by immune effector cells, represented by the term $\kappa CI$. The parameter $\kappa$ characterises the efficiency with which productively infected cells are eliminated.

Equation~\eqref{eq:model_V} describes the dynamics of free virions. New virions are produced by productively infected cells at rate $p$. The function $\varepsilon_{\mathrm{prod}}(t)\in[0,1]$ represents the efficacy of therapy in suppressing the production of new viral particles. Natural viral clearance is represented by the term $cV$, whereas the additional immune-mediated reduction in viral load is represented by $\varphi CV$.

Equation~\eqref{eq:model_C} governs the dynamics of immune effector cells. The term $s_C$ represents their basal influx. The saturating term $\alpha TC/(TC+h)$ describes stimulation of the effector response as a function of both the CD4$^+$ T-cell level and the current effector-cell concentration, with $h$ determining the saturation scale of this process. Natural loss of effector cells is represented by $d_CC$. The term $\eta_C C I/(I+q)$ describes functional exhaustion of the effector response under sustained antigenic stimulation, whereas $q$ specifies the characteristic infected-cell level at which this effect becomes pronounced.

Antiretroviral therapy is represented by two time-dependent efficacy functions, $\varepsilon_{\mathrm{inf}}(t)$ and $\varepsilon_{\mathrm{prod}}(t)$. The former reduces the rate at which new CD4$^+$ T cells become infected, whereas the latter suppresses the production of new virions. In the absence of treatment, we set
\begin{equation}
\varepsilon_{\mathrm{inf}}(t)
=
\varepsilon_{\mathrm{prod}}(t)
=
0.
\label{eq:no-therapy}
\end{equation}
The specific temporal profiles of these treatment efficacy functions used in the computational experiments are described in Subsection~\ref{subsec:numerical-methods}.

\subsection{Numerical Methods and Simulation Setup}
\label{subsec:numerical-methods}

We numerically solved system~\eqref{eq:model} as an initial-value problem
\begin{equation}
\frac{d\mathbf{y}}{dt}
=
\mathbf{f}(t,\mathbf{y}),
\qquad
\mathbf{y}(0)
=
\mathbf{y}_0,
\label{eq:cauchy-problem}
\end{equation}
where the state vector is defined as
\begin{equation}
\mathbf{y}(t)
=
\bigl(T(t),L(t),I(t),V(t),C(t)\bigr)^{\mathsf{T}}.
\label{eq:state-vector}
\end{equation}

System~\eqref{eq:model} is a nonlinear system of ordinary differential equations in which individual processes operate on substantially different timescales. In particular, the dynamics of free virions can be considerably faster than those of CD4$^+$ T cells, the latent reservoir, and the effector immune response. We therefore integrated the system using the adaptive LSODA algorithm, which switches automatically between Adams methods in non-stiff regions and backward differentiation formula (BDF) methods when stiffness is detected. This approach avoids the need to prescribe a single integration scheme for the entire simulation interval and is particularly suitable for representing rapid transient dynamics, treatment initiation, and subsequent changes in therapeutic efficacy.

We computed the solution over the interval $[0,T_{\mathrm{end}}]$ and recorded the state variables at predefined time points. These values were then used to construct temporal profiles and compare the therapeutic scenarios.

We assumed that no latently or productively infected cells were present initially:
\begin{equation}
L(0)=0,
\qquad
I(0)=0.
\label{eq:initial-infected}
\end{equation}
The initial concentrations of uninfected CD4$^+$ T cells and immune effector cells were specified as $T_0$ and $C_0$, respectively:
\begin{equation}
T(0)=T_0,
\qquad
C(0)=C_0.
\label{eq:initial-immune}
\end{equation}
The initial viral load was assigned a positive value,
\begin{equation}
V(0)=V_0.
\label{eq:initial-virus}
\end{equation}
In the baseline computational experiments, $V_0=1000$ was used. Thus,
\begin{equation}
\mathbf{y}_0
=
\bigl(T_0,0,0,V_0,C_0\bigr)^{\mathsf{T}}.
\label{eq:initial-vector}
\end{equation}

We selected parameter ranges based on published experimental estimates and parameterisations from related HIV models. For parameters lacking well-established experimental estimates, broader intervals were adopted to facilitate computational exploration. The principal parameter ranges are summarised in Table~\ref{tab:parameters}.

\begin{table}[tbp]
\centering
\caption{Model parameters and value ranges used in the numerical experiments
\cite{Perelson1993,Ho1995,Wei1995,Perelson1996,Perelson1997,
PerelsonNelson1999,RongPerelson2009,Conway2015}.}
\label{tab:parameters}

\small
\renewcommand{\arraystretch}{1.15}

\begin{tabularx}{\textwidth}{c X l}
\hline
Parameter & Biological interpretation & Range \\
\hline

$\lambda$
& influx of CD4$^+$ T cells
& $5$--$15$ $\mathrm{cells}/(\mu\mathrm{L}\cdot\mathrm{day})$ \\

$r$
& rate of homeostatic proliferation of CD4$^+$ T cells
& $0.05$--$0.2$ $\mathrm{day}^{-1}$ \\

$T_{\max}$
& effective carrying capacity of the CD4$^+$ T-cell population
& $500$--$1500$ $\mathrm{cells}/\mu\mathrm{L}$ \\

$d_T$
& natural death rate of CD4$^+$ T cells
& $5\times10^{-3}$--$2\times10^{-2}$ $\mathrm{day}^{-1}$ \\

$\beta$
& infection rate coefficient for CD4$^+$ T cells by free virions
& $10^{-8}$--$10^{-6}$
$\mu\mathrm{L}/(\mathrm{virion}\cdot\mathrm{day})$ \\

$\rho$
& fraction of newly infected cells entering the latent state
& $10^{-6}$--$2\times10^{-2}$ \\

$a$
& reactivation rate of latently infected cells
& $10^{-4}$--$10^{-2}$ $\mathrm{day}^{-1}$ \\

$\delta_L$
& natural death rate of latently infected cells
& $10^{-4}$--$10^{-2}$ $\mathrm{day}^{-1}$ \\

$\delta_I$
& natural death rate of productively infected cells
& $0.3$--$1.0$ $\mathrm{day}^{-1}$ \\

$p$
& rate of virion production by a productively infected cell
& $50$--$1000$ $\mathrm{virions}/(\mathrm{cell}\cdot\mathrm{day})$ \\

$c$
& natural clearance rate of free virions
& $1$--$5$ $\mathrm{day}^{-1}$ \\

$\varphi$
& coefficient of immune-mediated neutralisation of free virus
& $10^{-6}$--$10^{-4}$
$\mu\mathrm{L}/(\mathrm{cell}\cdot\mathrm{day})$ \\

$\kappa$
& coefficient of cytotoxic killing of productively infected cells
& $10^{-6}$--$10^{-4}$
$\mu\mathrm{L}/(\mathrm{cell}\cdot\mathrm{day})$ \\

$s_C$
& basal influx of immune effector cells
& $0.1$--$2$ $\mathrm{cells}/(\mu\mathrm{L}\cdot\mathrm{day})$ \\

$\alpha$
& maximum rate of effector immune-response stimulation
& $0.1$--$10$ $\mathrm{cells}/(\mu\mathrm{L}\cdot\mathrm{day})$ \\

$h$
& saturation parameter for immune stimulation
& $1$--$50$ $(\mathrm{cells}/\mu\mathrm{L})^2$ \\

$d_C$
& natural death rate of immune effector cells
& $0.05$--$0.2$ $\mathrm{day}^{-1}$ \\

$\eta_C$
& maximum rate of functional exhaustion of immune effector cells
& $10^{-3}$--$10^{-1}$ $\mathrm{day}^{-1}$ \\

$q$
& characteristic infected-cell level for saturation of the exhaustion effect
& $1$--$50$ $\mathrm{cells}/\mu\mathrm{L}$ \\

\hline
\end{tabularx}
\end{table}

Unless stated otherwise, the same parameter set and initial conditions were used when comparing therapeutic scenarios, with only the temporal profile of the therapeutic intervention being varied. This design allows differences between the resulting trajectories to be attributed directly to the treatment regimen under consideration.

Several treatment scenarios were considered in the computational analysis. In the baseline scenario, no therapy was administered throughout the entire simulation interval, as specified by condition~\eqref{eq:no-therapy}.

In the second scenario, ART was initiated at a prescribed time $t_0$. Because suppression of cellular infection and inhibition of viral production are represented separately in the model, the corresponding efficacy functions were defined as
\begin{equation}
\varepsilon_{\mathrm{inf}}(t)
=
\varepsilon_{\mathrm{inf},0}H(t-t_0),
\qquad
\varepsilon_{\mathrm{prod}}(t)
=
\varepsilon_{\mathrm{prod},0}H(t-t_0),
\label{eq:therapy-step}
\end{equation}
where $H$ denotes the Heaviside function, and
$\varepsilon_{\mathrm{inf},0}$ and
$\varepsilon_{\mathrm{prod},0}$ are the fixed efficacy levels following treatment initiation.

The third scenario represented periodic or irregular therapeutic exposure. In the general case, the efficacy functions were defined as
\begin{equation}
\varepsilon_{\mathrm{inf}}(t)
=
a_{\mathrm{inf}}(t)\varepsilon_{\mathrm{inf},0},
\qquad
\varepsilon_{\mathrm{prod}}(t)
=
a_{\mathrm{prod}}(t)\varepsilon_{\mathrm{prod},0},
\label{eq:therapy-periodic}
\end{equation}
where $a_{\mathrm{inf}}(t)\in[0,1]$ and
$a_{\mathrm{prod}}(t)\in[0,1]$ specify the temporal profiles of the therapeutic intervention. As a special case, the same temporal profile may be applied to both mechanisms of treatment action.

The fourth scenario accounted for a gradual decline in treatment efficacy associated with the development of drug resistance. Following treatment initiation at time $t_0$, exponentially decaying efficacy functions were used:
\begin{equation}
\varepsilon_{\mathrm{inf}}(t)
=
\varepsilon_{\mathrm{inf},0}
e^{-\gamma_{\mathrm{inf}}(t-t_0)}
H(t-t_0),
\qquad
\varepsilon_{\mathrm{prod}}(t)
=
\varepsilon_{\mathrm{prod},0}
e^{-\gamma_{\mathrm{prod}}(t-t_0)}
H(t-t_0),
\label{eq:therapy-resistance}
\end{equation}
where $\gamma_{\mathrm{inf}}>0$ and $\gamma_{\mathrm{prod}}>0$ characterise the rates at which the efficacy of the corresponding therapeutic mechanisms declines. If both mechanisms are assumed to follow the same decay law, one may set
$\gamma_{\mathrm{inf}}=\gamma_{\mathrm{prod}}=\gamma$.

The scenarios were compared on the basis of the dynamics of all model variables, $T(t)$, $L(t)$, $I(t)$, $V(t)$, and $C(t)$. The principal quantitative measures included the peak viral load, the minimum CD4$^+$ T-cell concentration, and the values of the state variables at the end of the simulation interval. The transient dynamics and long-term dynamical regimes were also examined, including infection suppression, persistence of a non-zero viral load, and viral rebound as treatment efficacy declined.

\section{Mathematical Analysis of the Model}\label{sec3}

This section examines the main mathematical properties of system
\eqref{eq:model} and relates the analytical threshold conditions to the
numerical dynamics of the model. We first establish positivity and boundedness
of the solutions. We then derive the infection-free equilibrium, the basic
reproduction number, and the corresponding local stability condition.
Finally, we analyse how antiretroviral therapy and viral latency modify the
infection threshold and derive a time-dependent criterion for loss of viral
control when treatment efficacy decreases.

Throughout this section, all model parameters are assumed to be positive,
with
$$
0\leq \rho \leq 1,\qquad
0\leq \varepsilon_{\mathrm{inf}}(t)\leq 1,\qquad
0\leq \varepsilon_{\mathrm{prod}}(t)\leq 1.
$$
For the autonomous analysis, the treatment efficacies are first assumed to be
constant,
$$
\varepsilon_{\mathrm{inf}}(t)=\varepsilon_{\mathrm{inf}},\qquad
\varepsilon_{\mathrm{prod}}(t)=\varepsilon_{\mathrm{prod}},
$$
and we introduce
$$
\theta_{\mathrm{inf}}=1-\varepsilon_{\mathrm{inf}},\qquad
\theta_{\mathrm{prod}}=1-\varepsilon_{\mathrm{prod}}.
$$

\subsection{Well-Posedness and Boundedness}\label{subsec:well-posedness}

The biological interpretation of system \eqref{eq:model} requires all state
variables to remain non-negative for non-negative initial data. Let
$$
\Omega_{+}=
\left\{(T,L,I,V,C)\in\mathbb{R}_{+}^{5}\right\}.
$$

\textbf{Proposition 1.}
\textit{For every initial condition
$(T(0),L(0),I(0),V(0),C(0))\in\Omega_{+}$, the corresponding solution of
system \eqref{eq:model} remains in $\Omega_{+}$ for all $t>0$ for which the
solution exists.}

\textit{Proof.}
On the boundary of the non-negative orthant,
\begin{align}
\left.\frac{dT}{dt}\right|_{T=0} &= \lambda>0,\\
\left.\frac{dL}{dt}\right|_{L=0}
&=\rho\theta_{\mathrm{inf}}\beta VT\geq0,\\
\left.\frac{dI}{dt}\right|_{I=0}
&=(1-\rho)\theta_{\mathrm{inf}}\beta VT+aL\geq0,\\
\left.\frac{dV}{dt}\right|_{V=0}
&=\theta_{\mathrm{prod}}pI\geq0,\\
\left.\frac{dC}{dt}\right|_{C=0} &= s_C>0.
\end{align}
Hence, the vector field points inward or is tangent to the boundary of
$\Omega_{+}$, so the non-negative orthant is positively invariant.
\hfill$\square$

The right-hand side of \eqref{eq:model} is locally Lipschitz continuous with
respect to the state variables. Therefore, a unique local solution exists
for every non-negative initial condition.

To establish boundedness, from \eqref{eq:model_T} we obtain
$$
\frac{dT}{dt}
\leq
\lambda+(r-d_T)T-\frac{r}{T_{\max}}T^2.
$$
The positive equilibrium of the corresponding scalar equation is
\begin{equation}
\overline{T}
=
\frac{T_{\max}}{2r}
\left[
(r-d_T)
+
\sqrt{(r-d_T)^2+\frac{4r\lambda}{T_{\max}}}
\right].
\label{eq:Tbar}
\end{equation}
Consequently,
$$
T(t)\leq M_T=\max\{T(0),\overline{T}\}.
$$

Next, let
$$
N(t)=T(t)+L(t)+I(t),
\qquad
d_*=\min\{d_T,\delta_L,\delta_I\}.
$$
Adding Eqs.~\eqref{eq:model_T}--\eqref{eq:model_I} gives
$$
\frac{dN}{dt}
=
\lambda
+rT\left(1-\frac{N}{T_{\max}}\right)
-d_TT-\delta_LL-\delta_II-\kappa CI.
$$
For $N\geq T_{\max}$,
$$
\frac{dN}{dt}\leq\lambda-d_*N,
$$
and therefore
$$
N(t)\leq
M_N=
\max\left\{
N(0),T_{\max},\frac{\lambda}{d_*}
\right\}.
$$
Since $I(t)\leq M_N$,
$$
\frac{dV}{dt}\leq pM_N-cV,
\qquad
V(t)\leq
M_V=
\max\left\{
V(0),\frac{pM_N}{c}
\right\}.
$$
Finally,
$$
0\leq\frac{\alpha TC}{TC+h}\leq\alpha,
$$
so that
$$
\frac{dC}{dt}\leq s_C+\alpha-d_CC,
\qquad
C(t)\leq
M_C=
\max\left\{
C(0),\frac{s_C+\alpha}{d_C}
\right\}.
$$
Thus, every solution starting from non-negative initial data remains
non-negative and bounded.

\subsection{Infection-Free Equilibrium and Invasion Threshold}
\label{subsec:infection-threshold}

At the infection-free equilibrium,
$$
L^*=I^*=V^*=0.
$$
The equilibrium concentration of uninfected CD4$^+$ T cells is given by
\eqref{eq:Tbar}. For $I=0$, the immune-response equilibrium satisfies
$$
s_C+\frac{\alpha\overline{T}C}{\overline{T}C+h}-d_CC=0.
$$
Introducing
$$
B_C=\overline{T}(s_C+\alpha)-d_Ch,
$$
the biologically relevant positive root is
\begin{equation}
\overline{C}
=
\frac{
B_C+
\sqrt{B_C^2+4d_C\overline{T}s_Ch}
}{
2d_C\overline{T}
}.
\label{eq:Cbar}
\end{equation}
Hence,
\begin{equation}
E_0=(\overline{T},0,0,0,\overline{C}).
\label{eq:DFE}
\end{equation}

To characterise viral invasion, consider the infected compartments
$(L,I,V)$ and define
$$
D_L=a+\delta_L,\qquad
D_I=\delta_I+\kappa\overline{C},\qquad
D_V=c+\varphi\overline{C}.
$$
A newly infected cell becomes productively infected immediately with
probability $1-\rho$. If it enters the latent compartment, activation occurs
before cell loss with probability $a/(a+\delta_L)$. Thus,
\begin{equation}
\xi_L
=
(1-\rho)+\rho\frac{a}{a+\delta_L}
=
1-\frac{\rho\delta_L}{a+\delta_L}.
\label{eq:xi-latency}
\end{equation}

The basic reproduction number is therefore
\begin{equation}
\mathcal{R}_0
=
\frac{
\theta_{\mathrm{inf}}
\theta_{\mathrm{prod}}
\beta p\overline{T}
}{
(\delta_I+\kappa\overline{C})
(c+\varphi\overline{C})
}
\left[
(1-\rho)+
\rho\frac{a}{a+\delta_L}
\right].
\label{eq:R0}
\end{equation}
It can be decomposed as
$$
\mathcal{R}_0
=
\mathcal{R}_{\mathrm{prod}}
+
\mathcal{R}_{\mathrm{lat}},
$$
where
\begin{align}
\mathcal{R}_{\mathrm{prod}}
&=
\frac{
\theta_{\mathrm{inf}}\theta_{\mathrm{prod}}
\beta p\overline{T}
}{
D_ID_V
}(1-\rho),
\\
\mathcal{R}_{\mathrm{lat}}
&=
\frac{
\theta_{\mathrm{inf}}\theta_{\mathrm{prod}}
\beta p\overline{T}
}{
D_ID_V
}
\rho\frac{a}{a+\delta_L}.
\end{align}

Linearisation of the infected subsystem about $E_0$ yields
\begin{equation}
J_{\mathrm{inf}}
=
\begin{pmatrix}
-D_L & 0 &
\rho\theta_{\mathrm{inf}}\beta\overline{T}
\\
a & -D_I &
(1-\rho)\theta_{\mathrm{inf}}\beta\overline{T}
\\
0 & \theta_{\mathrm{prod}}p & -D_V
\end{pmatrix}.
\label{eq:Jinfected}
\end{equation}
The next-generation matrix associated with \eqref{eq:Jinfected} leads to
\eqref{eq:R0}. The remaining eigenvalues associated with the uninfected
subsystem are negative at $E_0$. Hence:

\textbf{Theorem 1.}
\textit{The infection-free equilibrium $E_0$ is locally asymptotically
stable if $\mathcal{R}_0<1$ and unstable if $\mathcal{R}_0>1$.}

Thus, $\mathcal{R}_0=1$ defines the principal invasion threshold of the
model.

\subsection{Endemic Equilibrium and Baseline Dynamics}
\label{subsec:endemic-dynamics}

Let
$$
E_*=(T^*,L^*,I^*,V^*,C^*)
$$
denote an endemic equilibrium with $I^*>0$ and $V^*>0$. From the equilibrium
conditions for $L$ and $V$,
\begin{align}
L^*
&=
\frac{
\rho\theta_{\mathrm{inf}}\beta V^*T^*
}{
a+\delta_L
},
\label{eq:Lstar}
\\
V^*
&=
\frac{
\theta_{\mathrm{prod}}pI^*
}{
c+\varphi C^*
}.
\label{eq:Vstar}
\end{align}
Substitution into the equilibrium equation for $I$ gives
\begin{equation}
1
=
\frac{
\theta_{\mathrm{inf}}
\theta_{\mathrm{prod}}
\beta pT^*
}{
(\delta_I+\kappa C^*)
(c+\varphi C^*)
}
\left[
(1-\rho)+
\rho\frac{a}{a+\delta_L}
\right].
\label{eq:endemic-threshold}
\end{equation}
Accordingly, defining
\begin{equation}
\mathcal{R}(T,C)
=
\frac{
\theta_{\mathrm{inf}}
\theta_{\mathrm{prod}}
\beta pT
}{
(\delta_I+\kappa C)
(c+\varphi C)
}
\xi_L,
\label{eq:Rstate}
\end{equation}
every positive endemic equilibrium satisfies
\begin{equation}
\mathcal{R}(T^*,C^*)=1.
\label{eq:endemic-balance}
\end{equation}

Because of the nonlinear immune-stimulation and exhaustion terms, a closed
analytical expression for $E_*$ is not available in general. Its coordinates
are therefore obtained numerically.

\begin{figure}[t]
\centering
\includegraphics[width=0.92\textwidth]{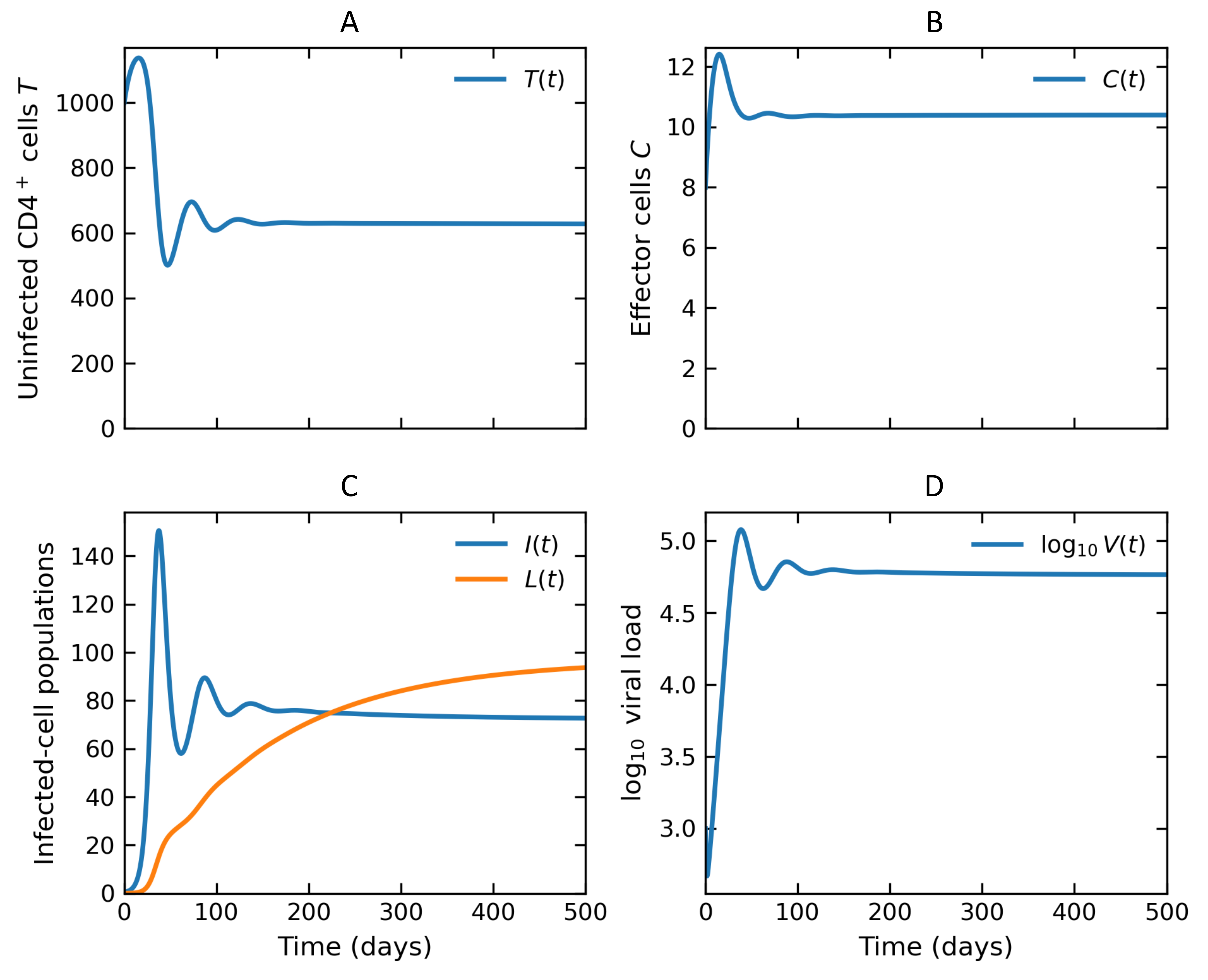}
\caption{Baseline within-host HIV dynamics in the absence of antiretroviral
therapy. (A) Uninfected CD4$^+$ T cells $T(t)$; (B) effector cells $C(t)$;
(C) productively infected cells $I(t)$ and latently infected cells $L(t)$;
(D) viral load $V(t)$ shown on a $\log_{10}$ scale.}
\label{fig:baseline-dynamics}
\end{figure}

Figure~\ref{fig:baseline-dynamics} illustrates the baseline dynamics in the
absence of ART. The solution remains bounded and approaches a persistent
infection state after an initial transient. The early viral increase is
accompanied by a marked reduction in the uninfected CD4$^+$ T-cell population
and a transient rise in productively infected cells. The latent compartment
accumulates more slowly and persists on a longer time scale, while the
effector-cell response approaches a non-zero stationary level.

\subsection{Effects of Antiretroviral Therapy and Viral Latency}
\label{subsec:art-latency}

Let
$$
\mathcal{R}_0^{(0)}
=
\frac{
\beta p\overline{T}
}{
(\delta_I+\kappa\overline{C})
(c+\varphi\overline{C})
}
\xi_L
$$
denote the reproduction number in the absence of treatment. Under constant
therapy,
\begin{equation}
\mathcal{R}_0^{(\mathrm{ART})}
=
(1-\varepsilon_{\mathrm{inf}})
(1-\varepsilon_{\mathrm{prod}})
\mathcal{R}_0^{(0)}.
\label{eq:RART}
\end{equation}
Hence, viral invasion is suppressed whenever
\begin{equation}
(1-\varepsilon_{\mathrm{inf}})
(1-\varepsilon_{\mathrm{prod}})
<
\frac{1}{\mathcal{R}_0^{(0)}}.
\label{eq:therapy-frontier}
\end{equation}
If both mechanisms have the same efficacy,
$\varepsilon_{\mathrm{inf}}=\varepsilon_{\mathrm{prod}}=\varepsilon$, the
critical efficacy is
\begin{equation}
\varepsilon
>
\varepsilon_{\mathrm{crit}}
=
1-\frac{1}{\sqrt{\mathcal{R}_0^{(0)}}}.
\label{eq:epsiloncrit}
\end{equation}

The effect of latency is contained in
$$
\xi_L=1-\frac{\rho\delta_L}{a+\delta_L}.
$$
Its derivatives are
\begin{align}
\frac{\partial\xi_L}{\partial\rho}
&=
-\frac{\delta_L}{a+\delta_L}<0,
\\
\frac{\partial\xi_L}{\partial a}
&=
\frac{\rho\delta_L}{(a+\delta_L)^2}>0,
\\
\frac{\partial\xi_L}{\partial\delta_L}
&=
-\frac{\rho a}{(a+\delta_L)^2}<0.
\end{align}
Thus, within the present model, increasing the fraction of newly infected
cells entering latency reduces the short-term invasion potential, whereas
increasing the reactivation rate increases it. This does not imply that
latency is protective in the long term: transfer into the latent compartment
provides a persistent source of cells that may later reactivate.

\begin{figure}[t]
\centering
\includegraphics[width=\textwidth]{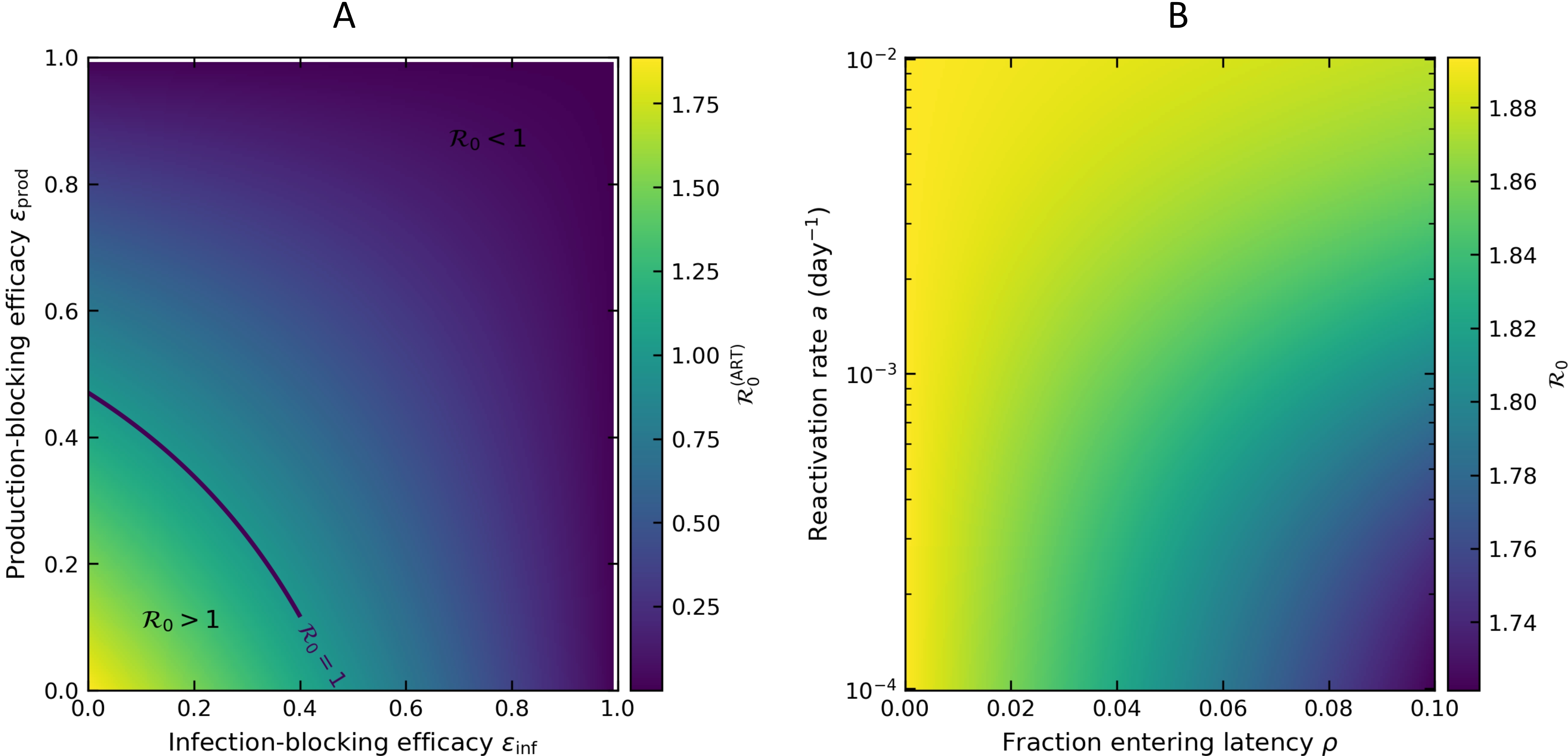}
\caption{Threshold landscapes predicted by the model. (A) Effective
reproduction number under constant ART as a function of infection-blocking
efficacy $\varepsilon_{\mathrm{inf}}$ and production-blocking efficacy
$\varepsilon_{\mathrm{prod}}$; the contour
$\mathcal{R}_0^{(\mathrm{ART})}=1$ separates the persistence and suppression
regions. (B) Dependence of the basic reproduction number on the fraction
$\rho$ of newly infected cells entering latency and the latent-cell
reactivation rate $a$.}
\label{fig:threshold-landscapes}
\end{figure}

These two effects are summarised in Fig.~\ref{fig:threshold-landscapes}.
Panel A shows the therapeutic threshold in the
$(\varepsilon_{\mathrm{inf}},\varepsilon_{\mathrm{prod}})$ plane, with the
contour $\mathcal{R}_0^{(\mathrm{ART})}=1$ separating parameter combinations
that allow invasion from those that suppress it. Panel B shows the dependence
of $\mathcal{R}_0$ on the latent fraction $\rho$ and the reactivation rate
$a$. For the baseline parameter set and the investigated latency range,
variation in $(\rho,a)$ changes the magnitude of $\mathcal{R}_0$ but does not
by itself move the system below the invasion threshold.

\subsection{Time-Dependent Therapy and Loss of Viral Control}
\label{subsec:time-dependent-therapy}

When treatment efficacy varies in time, the system is non-autonomous and a
single constant reproduction number is no longer sufficient to describe the
instantaneous infection pressure. We therefore define
\begin{equation}
\mathcal{R}_{\mathrm{eff}}(t)
=
\frac{
[1-\varepsilon_{\mathrm{inf}}(t)]
[1-\varepsilon_{\mathrm{prod}}(t)]
\beta pT(t)
}{
[\delta_I+\kappa C(t)]
[c+\varphi C(t)]
}
\xi_L.
\label{eq:Reff}
\end{equation}
This quantity is used here as a local-in-time threshold indicator rather than
as a formal reproduction number for the general non-autonomous system.
Intervals with $\mathcal{R}_{\mathrm{eff}}(t)<1$ correspond to conditions
unfavourable for viral amplification, whereas
$\mathcal{R}_{\mathrm{eff}}(t)>1$ indicates that the current biological and
therapeutic state permits renewed viral growth.

To represent progressive loss of treatment efficacy, suppose that after
therapy initiation at $t=t_0$,
$$
\varepsilon_{\mathrm{inf}}(t)
=
\varepsilon_{\mathrm{prod}}(t)
=
\varepsilon_0e^{-\gamma(t-t_0)},
\qquad t\geq t_0.
$$
Using the infection-free equilibrium as a reference state gives
\begin{equation}
\mathcal{R}_{\mathrm{ART}}(t)
=
\mathcal{R}_0^{(0)}
\left[
1-\varepsilon_0e^{-\gamma(t-t_0)}
\right]^2.
\label{eq:Rresistance}
\end{equation}
If $\mathcal{R}_0^{(0)}>1$ and the initial efficacy satisfies
$$
\varepsilon_0
>
1-\frac{1}{\sqrt{\mathcal{R}_0^{(0)}}},
$$
the time at which the suppression threshold is lost follows from
$\mathcal{R}_{\mathrm{ART}}(t_{\mathrm{loss}})=1$:
\begin{equation}
t_{\mathrm{loss}}
=
t_0
+
\frac{1}{\gamma}
\ln
\left[
\frac{\varepsilon_0}
{
1-1/\sqrt{\mathcal{R}_0^{(0)}}
}
\right].
\label{eq:tloss}
\end{equation}

\begin{figure}[t]
\centering
\includegraphics[width=0.78\textwidth]{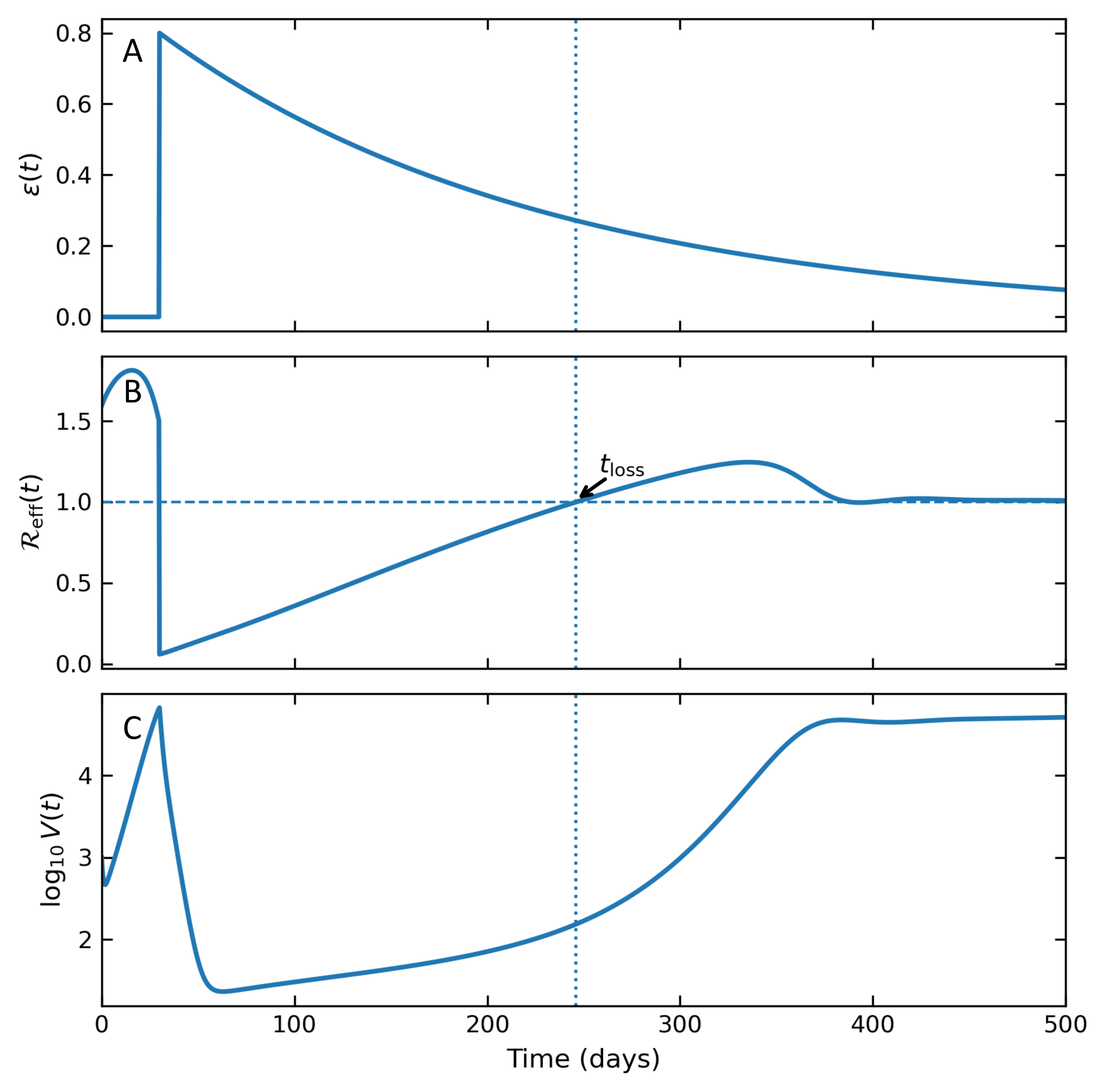}
\caption{Loss of viral control under declining ART efficacy. (A) Time-dependent
treatment efficacy $\varepsilon(t)$ after therapy initiation; (B) effective
reproduction potential $\mathcal{R}_{\mathrm{eff}}(t)$ with the horizontal
threshold $\mathcal{R}_{\mathrm{eff}}=1$ and the analytically predicted
threshold-loss time $t_{\mathrm{loss}}$; (C) corresponding viral load on a
$\log_{10}$ scale.}
\label{fig:loss-control}
\end{figure}

Figure~\ref{fig:loss-control} links this analytical prediction with the full
model dynamics. After therapy initiation, viral replication is strongly
suppressed and $\mathcal{R}_{\mathrm{eff}}(t)$ falls below unity. As treatment
efficacy gradually decreases, $\mathcal{R}_{\mathrm{eff}}(t)$ returns to the
threshold value at $t=t_{\mathrm{loss}}$. The subsequent increase in viral
load occurs after this threshold crossing, showing that loss of the
mathematical suppression condition precedes the pronounced viral rebound in
the simulated trajectory. Accordingly, $t_{\mathrm{loss}}$ should be
interpreted as a mechanistic threshold time, not as a clinical detection time.

Overall, the analysis identifies a common threshold structure underlying the
autonomous and time-dependent versions of the model. The basic reproduction
number determines local invasion of the infection-free state, ART modifies
this threshold multiplicatively through the two treatment mechanisms, and
declining treatment efficacy can be translated into an explicit
threshold-loss time. At the same time, the latency contribution separates
short-term invasion potential from long-term persistence through the latent
reservoir.

\section{Results of computational experiments}\label{sec4}

This section presents the results of computational experiments conducted to investigate characteristic regimes of HIV infection dynamics described by system~\eqref{eq:model}. Three principal scenarios were considered: the natural course of infection in the absence of treatment, initiation of antiretroviral therapy (ART) at a prescribed time point, and periodic variation in therapeutic efficacy.

In all experiments, the dynamics of five model variables were analysed: the concentration of uninfected CD4$^+$ T cells $T(t)$, latently infected cells $L(t)$, productively infected cells $I(t)$, free virions $V(t)$, and immune effector cells $C(t)$. This formulation enables simultaneous assessment of changes in viral load, the status of the target-cell population, formation of the latent reservoir, and the response of the effector arm of the immune system.

\subsection{HIV infection dynamics in the absence of antiretroviral therapy}
\label{subsec:no-treatment}

We first considered HIV dynamics in the absence of treatment. Throughout the entire simulation interval, we assumed
$$
\varepsilon_{\mathrm{inf}}(t)=
\varepsilon_{\mathrm{prod}}(t)=0.
$$
Accordingly, the rate of infection of susceptible CD4$^+$ T cells and the rate of production of new virions were determined solely by the intrinsic parameters of the system.

\begin{figure}[htbp]
    \centering
    \includegraphics[width=0.8\textwidth]{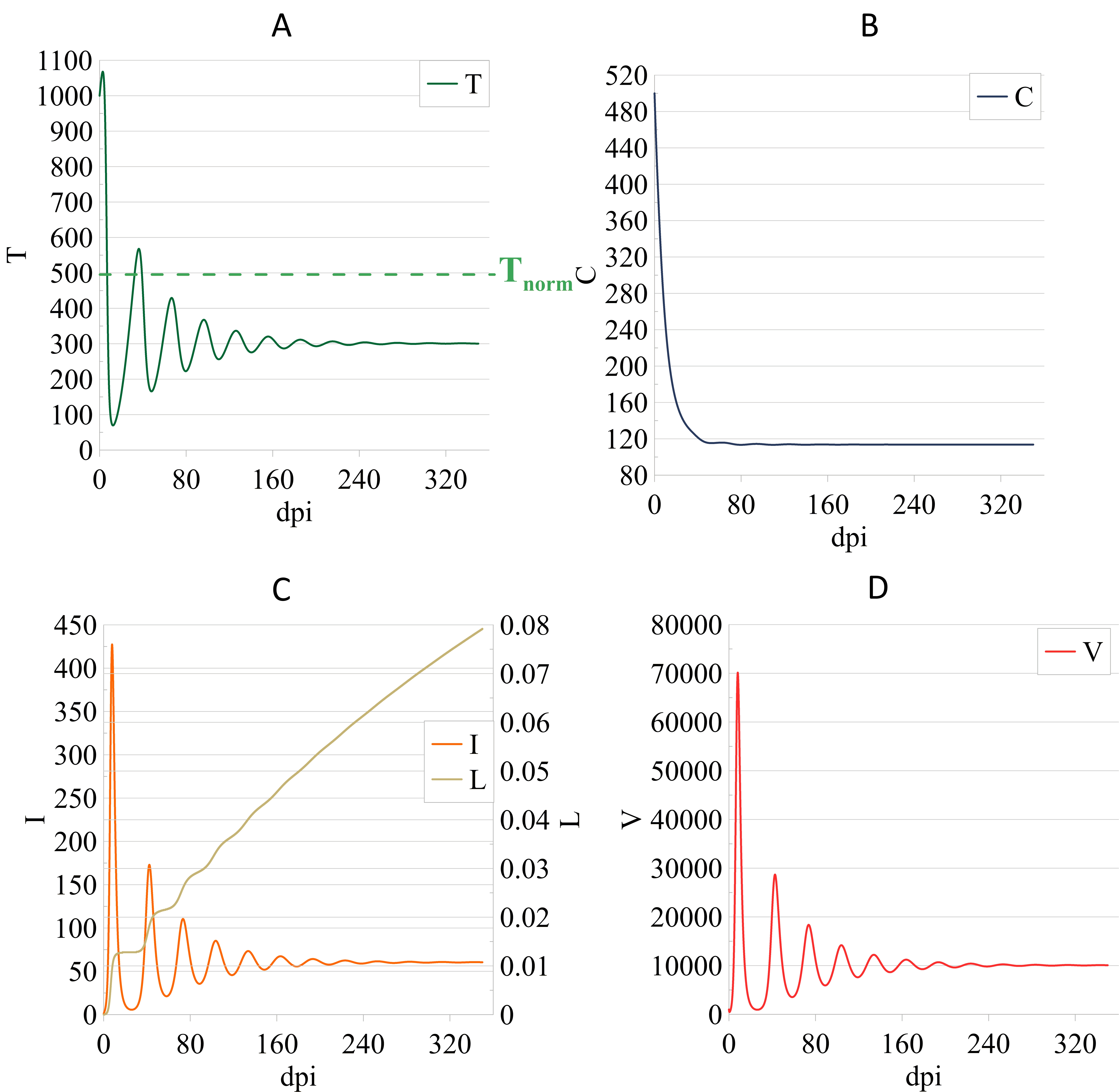}
    \caption{Dynamics of the model variables in the absence of antiretroviral
    therapy: (A) concentration of uninfected CD4$^+$ T cells $T(t)$;
    (B) concentration of immune effector cells $C(t)$;
    (C) concentrations of productively infected $I(t)$ and latently
    infected $L(t)$ cells; (D) concentration of free virions $V(t)$.}
    \label{fig:without-treatment}
\end{figure}

The results of this computational experiment are presented in Fig.~\ref{fig:without-treatment}. The simulation shows a pronounced initial transient. The concentration of free virions $V(t)$ increases rapidly and reaches an initial peak of approximately $7\times10^{4}$ in the model units used here (Fig.~\ref{fig:without-treatment}D). At the same time, the concentration of productively infected cells $I(t)$ rises sharply (Fig.~\ref{fig:without-treatment}C).

The increase in viral load is accompanied by a rapid decline in the number of uninfected CD4$^+$ T cells. Following the initial transient, $T(t)$ exhibits a series of damped oscillations and gradually approaches a steady-state level of approximately $300$ (Fig.~\ref{fig:without-treatment}A). This value remains below the reference level $T_{\mathrm{norm}}$ shown in the figure, corresponding to marked depletion of the target-cell population.

The dynamics of effector cells $C(t)$ are characterised by a rapid decline during the initial phase, followed by stabilisation at an approximately constant non-zero level (Fig.~\ref{fig:without-treatment}B). This behaviour results from the combined effects of basal effector-cell influx, immune stimulation, natural cell loss, and functional exhaustion under persistent antigenic stimulation.

The two infected-cell compartments exhibit markedly different timescales. Following a pronounced primary peak, the population of productively infected cells $I(t)$ undergoes a sequence of damped oscillations before approaching a non-zero steady-state level. By contrast, $L(t)$ changes much more slowly and increases gradually throughout the entire simulation interval (Fig.~\ref{fig:without-treatment}C). Thus, the model predicts a progressive accumulation of a latently infected cellular reservoir.

After the transient phase has subsided, the viral load likewise does not approach zero but instead stabilises at a level of approximately $10^{4}$ (Fig.~\ref{fig:without-treatment}D). Therefore, for the selected parameter set, the system approaches a regime of persistent infection in which $I(t)$, $L(t)$, and $V(t)$ remain non-zero. This finding is consistent with the mathematical analysis presented in Section~\ref{sec3}: when $\mathcal{R}_0>1$, the infection-free equilibrium loses stability and the virus can sustain long-term persistence within the host.

\subsection{Effect of antiretroviral therapy initiation on system dynamics}
\label{subsec:therapy-onset}

The second computational experiment investigated changes in system dynamics following the initiation of antiretroviral therapy. Under the selected scenario, no therapeutic intervention was applied before $t_0=100$ days, such that
$$
\varepsilon_{\mathrm{inf}}(t)=
\varepsilon_{\mathrm{prod}}(t)=0,
\qquad t<t_0.
$$
From $t=t_0$ onwards, the efficacy of suppressing infection of new cells was set to
$$
\varepsilon_{\mathrm{inf}}=0.6,
$$
whereas the efficacy of suppressing production of new virions was set to
$$
\varepsilon_{\mathrm{prod}}=0.5.
$$
Thus, treatment initiation was modelled as a stepwise change in the therapeutic efficacy coefficients.

\begin{figure}[htbp]
    \centering
    \includegraphics[width=0.8\textwidth]{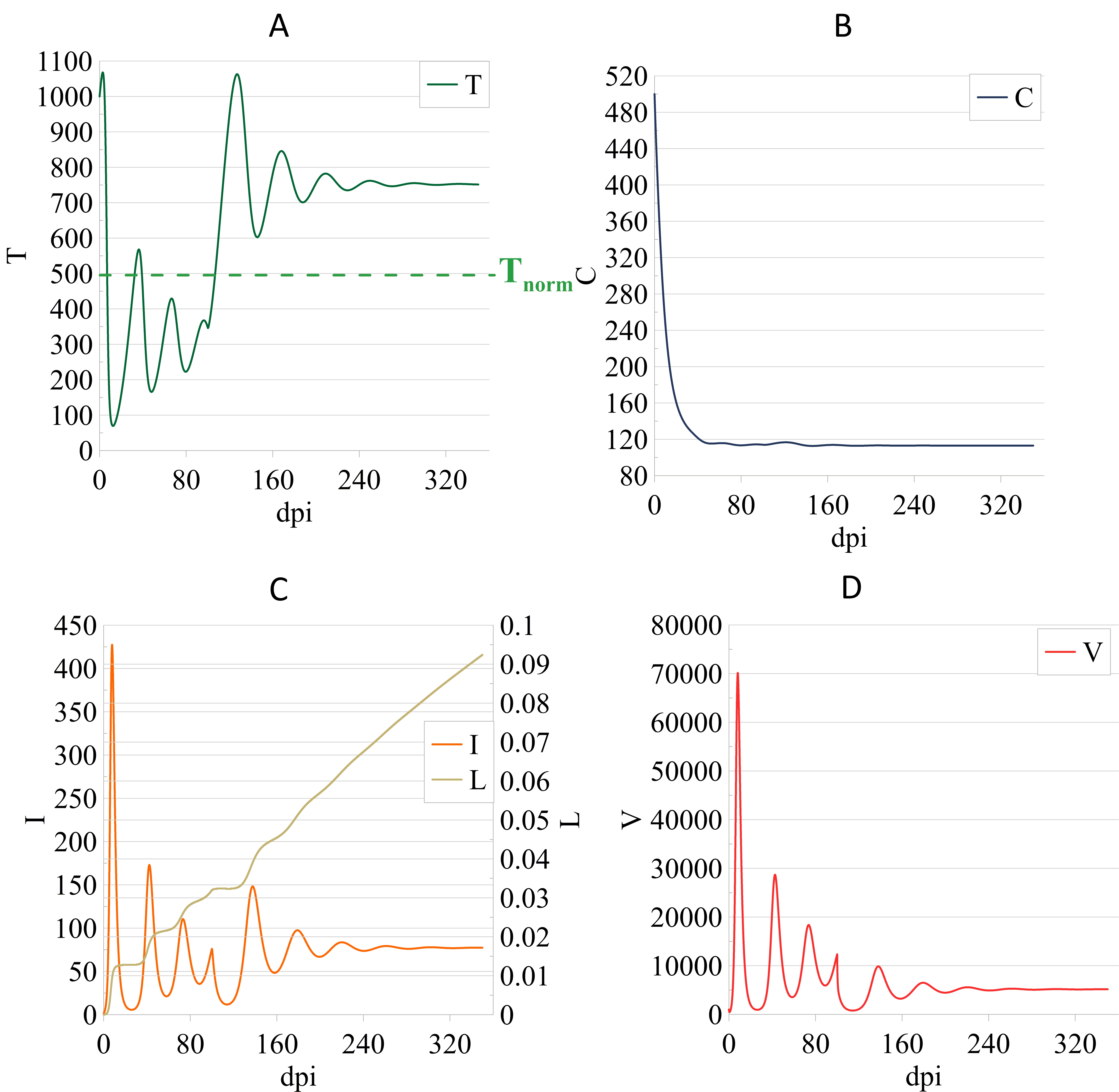}
    \caption{Dynamics of the model variables following initiation of antiretroviral
    therapy on day $100$ after infection
    ($\varepsilon_{\mathrm{inf}}=0.6$,
    $\varepsilon_{\mathrm{prod}}=0.5$):
    (A) concentration of uninfected CD4$^+$ T cells $T(t)$;
    (B) concentration of effector cells $C(t)$;
    (C) concentrations of productively infected $I(t)$ and latently
    infected $L(t)$ cells;
    (D) concentration of free virions $V(t)$.}
    \label{fig:therapy-onset}
\end{figure}

The results are presented in Fig.~\ref{fig:therapy-onset}. Before treatment initiation, the system dynamics are virtually identical to those observed for the untreated course of infection, with a pronounced increase in viral load, depletion of the uninfected CD4$^+$ T-cell population, and oscillatory dynamics of the infected-cell compartments.

ART initiation substantially changes the system dynamics. The most pronounced change is observed for $T(t)$ (Fig.~\ref{fig:therapy-onset}A). After a transient period, the concentration of uninfected CD4$^+$ T cells increases and stabilises at approximately $750$, substantially above the corresponding value in the untreated scenario. This result indicates that reducing the rate at which new cells become infected enables substantial recovery of the CD4$^+$ T-cell population.

At the same time, the steady-state viral load decreases (Fig.~\ref{fig:therapy-onset}D). In the absence of treatment, the late-time value of $V(t)$ is approximately $10^{4}$, whereas after ART initiation it falls to about $5\times10^{3}$ in the model units used here. However, the viral concentration does not decline to zero. Therefore, for the selected values of $\varepsilon_{\mathrm{inf}}$ and $\varepsilon_{\mathrm{prod}}$, therapy substantially suppresses viral replication but does not achieve complete elimination of infection.

The population of productively infected cells $I(t)$ also shifts following treatment initiation and, after several transient oscillations, approaches a new steady-state regime (Fig.~\ref{fig:therapy-onset}C). At the same time, latently infected cells $L(t)$ persist within the system. Their concentration changes much more slowly than those of $I(t)$ and $V(t)$, reflecting the long-lived nature of the latent compartment.

Following the rapid initial transient, $C(t)$ changes comparatively little and stabilises at a non-zero level (Fig.~\ref{fig:therapy-onset}B). Hence, under the therapeutic regimen considered here, the principal effects of ART are manifested primarily as a reduction in viral load and recovery of the uninfected CD4$^+$ T-cell population.

Overall, treatment initiation shifts the system towards a qualitatively more favourable regime compared with the untreated scenario. Nevertheless, non-zero values of $V(t)$, $I(t)$, and $L(t)$ indicate persistence of both active infection and the latent cellular reservoir.

\subsection{System dynamics under periodically varying treatment efficacy}
\label{subsec:periodic-therapy}

We next considered a scenario in which treatment efficacy varies over time. This formulation makes it possible to account for imperfect treatment adherence, periodic fluctuations in drug concentration, and other factors for which ART efficacy cannot reasonably be regarded as constant.

In accordance with the model formulation, the time dependence of therapeutic efficacy was specified as
$$
\varepsilon(t)=a(t)\varepsilon_0,
$$
where $a(t)\in[0,1]$ defines the temporal profile of drug action. A smooth function with alternating phases of increasing and decreasing efficacy was used to generate the periodic profile.

\begin{figure}[htbp]
    \centering
    \includegraphics[width=0.8\textwidth]{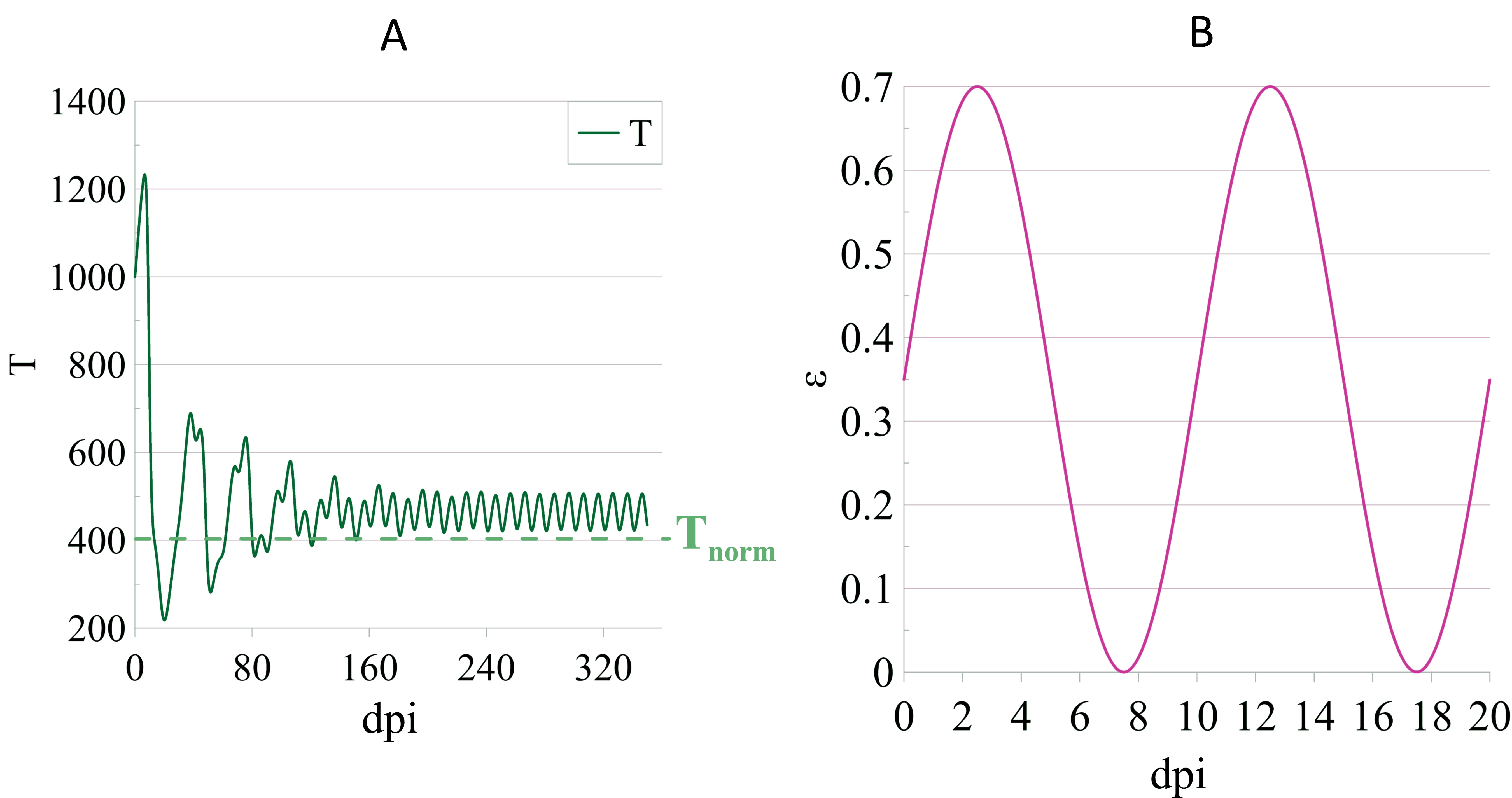}
    \caption{System dynamics under periodically varying antiretroviral therapy efficacy:
    (A) concentration of uninfected CD4$^+$ T cells $T(t)$;
    the dashed line indicates the reference level $T_{\mathrm{norm}}$;
    (B) temporal profile of therapeutic efficacy $\varepsilon(t)$.}
    \label{fig:periodic-therapy}
\end{figure}

Figure~\ref{fig:periodic-therapy}B shows the prescribed profile $\varepsilon(t)$. In the scenario considered, efficacy varies periodically from values close to zero to a maximum of approximately $0.7$. The corresponding dynamics of the CD4$^+$ T-cell population are shown in Fig.~\ref{fig:periodic-therapy}A.

During the initial phase, $T(t)$ exhibits large-amplitude oscillations arising from both the intrinsic transient dynamics of the system and the time variation in therapeutic efficacy. As time progresses, the contribution of the intrinsic transient diminishes; however, the oscillations do not disappear completely. At later times, the system approaches a regime of small periodic oscillations around an intermediate mean level.

Thus, even after the initial transient has subsided, periodic modulation of ART efficacy continues to influence the CD4$^+$ T-cell population. In contrast to the case of constant therapeutic efficacy, the system does not converge to a strictly stationary state but instead approaches a forced oscillatory regime.

These results demonstrate that disease dynamics are determined not only by the mean efficacy of treatment but also by its temporal distribution. Periods of reduced therapeutic efficacy create conditions for renewed intensification of the infectious process, whereas subsequent increases in ART efficacy promote recovery of the cell population. Explicit representation of the time dependence of $\varepsilon_{\mathrm{inf}}(t)$ and $\varepsilon_{\mathrm{prod}}(t)$ is therefore important when modelling irregular treatment regimens.

\section{Discussion}\label{sec5}

The model combines latent infection, cytotoxic immune control, and time-dependent ART within a single ODE framework. The analytical and numerical results reveal a common threshold structure. The condition $R_0<1$ determines the local stability of the infection-free state, while changes in infection rate, virion production, and treatment efficacy alter the position of the system relative to this threshold. The simulations show how these effects translate into different patterns of viral and cellular dynamics. This combination of threshold analysis and time-dependent simulations is consistent with current developments in mathematical modelling of HIV infection, where equilibrium and stability analyses are increasingly complemented by investigations of treatment-dependent and data-informed dynamics \cite{Perelson2013,Hill2018Treatment,Liyanage2024}.

In the absence of therapy, the system approaches a regime of persistent infection: following the initial transient, non-zero populations of productively infected cells and free virus are maintained, while the level of uninfected CD4$^+$ T cells remains reduced. The observed damped oscillations should be interpreted as intrinsic transient dynamics of the nonlinear system arising from interactions among viral replication, target-cell turnover, and the immune response, rather than as a direct quantitative reproduction of clinical oscillatory behaviour. The persistence of a non-zero long-term viral load is consistent with the classical view that, once the invasion threshold is exceeded, infection can be sustained through the continuous generation of newly infected cells \cite{Perelson2013,Hill2018Treatment}. At the same time, the effector compartment $C(t)$ provides partial immune control but, under the parameterisation considered here, does not result in elimination of infection. This behaviour is consistent with the well-established but limited role of the CD8$^+$ T-cell response in controlling chronic HIV infection \cite{Collins2020}.

Initiation of ART reduces viral load and promotes recovery of the uninfected CD4$^+$ T-cell population, in qualitative agreement with both clinical observations and classical models of viral dynamics \cite{START2015,Hill2018Treatment}. However, for the selected values of $\varepsilon_{\mathrm{inf}}$ and $\varepsilon_{\mathrm{prod}}$, the viral load remains non-zero. This result should not be interpreted as a quantitative prediction of the efficacy of contemporary combination ART. The therapeutic coefficients used in the present study are phenomenological parameters, and the computational experiment is intended to investigate mechanisms underlying transitions between dynamical regimes rather than to reproduce a specific pharmacological regimen.

Of particular importance is the difference in timescales between active infection and the latent compartment. Following treatment initiation, $V(t)$ and $I(t)$ respond relatively rapidly to changes in therapeutic efficacy, whereas $L(t)$ evolves much more slowly and persists even when active viral replication is suppressed. This behaviour reflects one of the central biological features of HIV infection: standard ART suppresses viral replication but does not eliminate the latent reservoir. The long-term stability of latently infected CD4$^+$ T cells and their capacity to act as a source of subsequent viral reactivation are well established experimentally and remain major barriers to HIV eradication \cite{Siliciano2003,RongPerelson2009,Rasi2025}. In mathematical models, reservoir size and the strength of immune control likewise exert a substantial influence on post-treatment dynamics and the possibility of sustained remission \cite{Conway2015,Vemparala2024}.

In the present model, however, latency is represented in an aggregated form. The latent compartment is generated from newly infected cells, depleted through cell loss, and contributes to the productively infected population through reactivation. The model does not account for clonal expansion of latently infected cells, heterogeneity among proviruses with respect to their capacity for reactivation, or anatomical compartmentalisation of the reservoir. Accordingly, the dynamics of $L(t)$ should be interpreted as a qualitative representation of latent-reservoir persistence rather than as a quantitative prediction of its absolute size or half-life \cite{RongPerelson2009,Rasi2025}.

The experiment with periodically varying treatment efficacy demonstrates that system dynamics depend not only on the mean level of therapeutic activity but also on its temporal distribution. After the initial transient has decayed, periodic modulation of $\varepsilon(t)$ sustains forced oscillations in the cell population and prevents convergence to a strictly stationary regime. This behaviour arises from the nonlinear structure of the model: intervals of reduced and increased efficacy do not necessarily compensate for one another under simple temporal averaging. Similar interactions between the temporal structure of treatment and characteristic timescales of viral dynamics have been considered in other mathematical models of periodic antiviral therapy \cite{Browne2020}. Nevertheless, the smooth profile $\varepsilon(t)$ employed here represents an idealised description of therapeutic exposure and should not be interpreted directly as the pharmacokinetic profile of a specific drug or as an exact model of missed doses. Addressing such questions would require explicit pharmacokinetic and pharmacodynamic components, because the relationship between treatment adherence and virological suppression depends strongly on the ART regimen considered \cite{Manalel2024}.

The threshold analysis further complements this interpretation. The derived expression for $\mathcal{R}_0^{(\mathrm{ART})}$ indicates that suppression of new-cell infection and inhibition of viral production jointly determine the position of the system relative to the threshold $\mathcal{R}_0=1$. Under time-dependent therapy, $\mathcal{R}_{\mathrm{eff}}(t)$ is used as a local indicator of whether the current conditions are favourable or unfavourable for renewed viral growth. In the scenario of progressively declining treatment efficacy, the time $t_{\mathrm{loss}}$ at which this quantity reaches unity precedes pronounced viral rebound. Thus, loss of the mathematical condition required for infection suppression and the onset of a readily observable increase in viral load represent distinct events. Importantly, $t_{\mathrm{loss}}$ should be regarded as a mechanistic characteristic of the model rather than as a prediction of the time at which virological failure would be detected clinically.

The exponential decline in treatment efficacy is likewise a phenomenological representation of drug resistance. In reality, resistance is determined by the emergence and selection of specific mutations, the genetic barrier of individual drugs, prior treatment exposure, and the level of drug exposure. The parameter $\gamma$ in the present model therefore characterises an aggregated rate of loss of therapeutic efficacy rather than the evolution of a specific resistant viral variant. Nevertheless, this approximation enables investigation of the dynamical consequences of progressively weakening therapeutic control, which remains relevant given the continuing clinical challenge posed by antiretroviral drug resistance \cite{Kagan2025}.

The principal limitation of the present study is the absence of calibration against individual longitudinal clinical data. The resulting trajectories should therefore be regarded as mechanistic scenarios rather than personalised predictions. Additional limitations include the aggregated representation of the effector immune response, the absence of spatial and stochastic structure, and the use of phenomenological ART efficacy functions. Of particular importance is the potentially limited identifiability of parameters associated with the immune subsystem, which is characteristic of comparable within-host models \cite{Liyanage2024}.

Further development of the model may include calibration against longitudinal viral-load and cellular data, analysis of structural and practical identifiability, global sensitivity analysis, and quantification of parameter uncertainty. Promising extensions include explicit representation of ART pharmacokinetics, clonal dynamics of the latent reservoir, emergence of resistant viral variants, and stochastic reactivation. Nevertheless, an important advantage of the proposed framework is its comparatively compact structure, which preserves analytical tractability while simultaneously accounting for active infection, latency, immune control, and time-dependent therapeutic intervention.

\section{Conclusion}\label{sec6}

We developed a five-compartment ODE model that combines productive and latent HIV infection, cytotoxic immune control, and time-dependent ART. The analysis establishes non-negativity and boundedness of the solutions and gives an invasion threshold that depends explicitly on the two treatment efficacy functions. For exponentially declining efficacy, the model also provides an analytical estimate of the threshold-crossing time, which precedes viral rebound in the simulations.

The numerical experiments show how ART initiation and temporal variation in treatment efficacy alter viral load, CD4$^+$ T-cell recovery, and persistence of the latent reservoir. Because the model has not been calibrated against longitudinal patient data, these trajectories should be interpreted as mechanistic scenarios rather than clinical predictions. The next step is to examine identifiability and calibrate the model, followed by sensitivity and uncertainty analyses and, where supported by data, more detailed descriptions of pharmacokinetics, resistance, and latent-reservoir dynamics.
\backmatter

\section*{Declarations}

\hspace*{\parindent}\textbf{Funding.} The work was supported by the Ministry of Science and Higher Education of the Russian Federation (project FZUU-2026-0005).

\textbf{Code availability.}
The source code supporting the mathematical model, numerical simulations, and computational experiments presented in this study is publicly available in the GitHub repository: \url{https://github.com/Blank369/hiv-model_app.git}.

\textbf{Conflict of interest.}
The authors declare that they have no conflict of interest.

\textbf{Author contributions.} 
M.P. contributed to conceptualization, methodology, mathematical analysis, supervision, and writing. Y.C. contributed to software development, numerical simulations, investigation, visualization, and writing--original draft. Both authors reviewed and approved the final manuscript.

\textbf{Data availability.}
No new data were created or analysed in this study.

\bigskip





\bibliography{sn-bibliography}

\end{document}